# Theoretical investigation of electronic, optical and thermoelectric properties of tellurium doped barium titanate (BTO) through modified Becke – Johnson exchange potential.

Meryem Ziati [1, *] and Hamid Ez-Zahraouy [1]

[1] Laboratory of Condensed Matter and Interdisciplinary Sciences department of physics, Faculty of Sciences, Mohammed V University in Rabat – Morocco.

*Corresponding author: meryem.ziati.um5@gmail.com

**A b s t r a c t:**

The stability, electronic structure, optical and thermoelectric properties of Te-doped $BaTiO_3$ are investigated by first-principal calculation based on the density functional theory and Boltzmann transport theory implemented in WIEN2K and BoltzTraP simulation program. This study is carried out by applying LDA + TB-mBJ potential. Formation energy of each doped structure is calculated to examine the stability and feasibility of the synthesis. Incorporating Te into $BaTiO_3$ efficiently reduces the electronic band gap and the level of band gap reduction can be controlled by varying the amount of dopant ($E_g$ = 2.752 eV for pure BTO and $E_g$ = 0.500 eV for Te doped-BTO: 8.3%). Hence, the absorption ability is improved in the visible light (380~790 nm). Our findings suggest that all the doped structures are significantly absorbent and productive with an optical absorption that exceeds $10^5$ cm$^{-1}$ in the visible range ($\lambda \sim 500$ nm). In addition, $BaTiO_3$ revealed a smaller dielectric constant at zero frequency ($\varepsilon_1(0)$ = 3.98) compared to Te doped-BTO: 8.3% ($\varepsilon_1(0)$ = 5.55), while the optical energy gap is reduced from 3.692 eV to 1.619 eV by growing Te concentration. Then, optical conductivity and Urbach's parameters are predicted. The transport properties are assessed as a function of temperature. It is found that the electrical conductivity is considerably enhanced with increasing Te concentration. Other thermoelectric properties such as Seebeck coefficient and figure of merit are also investigated. Our theoretical results can be useful for thermoelectric and visible-light photoelectrical device applications.

# 1. INDRODUCTION

The discovery of calcium titanate ($CaTiO_3$) in 1839 by a Russian mineralogist Perovski is regarded as the origin of perovskite, and materials with the same type of crystal structure as that of $CaTiO_3$ are known as the perovskite materials [1].

The perovskite structure is well known to provide excellent electrical and optical properties that make it suitable for a variety of electronic applications (infrared sensors, electromechanical transducers, and optical modulators). Most crucially, the electrical properties can be engineered via chemical or physical modifications to obtain insulating, semiconducting, and conducting features [2].

Perovskite type oxides received a lot of interest due to their wide range of valuable properties such as ferro-electricity, piezo-electricity, pyro-electricity, magnetism, and electro-optic effects [3]. The typical chemical formula of the perovskite oxides is $ABO_3$, where A and B designate two different cations. This structure is comprised of a large-sized 12-coordinated cations at the A-site and small-sized 6-coordinated cations at the B-site [4]. Furthermore, the ideal structure of perovskites is a cubic lattice with space group Pm3m, in which the B – O and the A – O distances in these perovskites are also roughly equal to $(a/2)$ and $(a/\sqrt{2})$ respectively, wherein "$a$" represents the lattice constant, and the distance A – O is about 40% larger than the B – O bond distance. If the ions are considered as rigid balls, all the ions in $ABO_3$ should be in a mutually tangent state and therefore the lattice parameter $a = \sqrt{2}\ (r_{A+}r_O) = 2\ (r_{B+}r_O)$ [5]. Normally, in real materials, the value of $\sqrt{2}\ (r_{A+}r_O)$ is not equal to that of $2\ (r_{B+}r_O)$ but they can always form a cubic perovskite structure. Most crucially, the tolerance factor $t = (r_{A+}r_O)/\sqrt{2}(r_{B+}r_O)$ given by Goldschmidt early in the 1920 [6] performs a primordial role in assessing the stability of perovskites and indicates that the cubic structure can be retained for $0.95 \leq t \leq 1.04$ [6].

Even though compounds have this ideal cubic structure, many oxides exhibit slightly distorted variants with smaller symmetry such as orthorhombic or rombohedral lattice (Khattak et Wang., 1979, PEROVSKITES AND GARNETS). As an example, the perovskites $LaCoO_3$ and $LaMnO_3$ present a rhombohedral deformation (Forni et al. 1996; Alifanti et al. 2003 [7]). Whereas, in the case of the $LaFeO_3$ compound, an orthorhombic distortion is observed (Marezio et al. 1971. [8]).

Barium titanate ($BaTiO_3$) is one of the most prominent perovskite crystals, which has been the subject of many experimental and theoretical investigations. As an example, Fuentes et al. [9] prepared $BaTiO_3$ powders using a sol-gel-hydrothermal process to study its effects at different times and calcination temperatures. Ali et al. [10] prepared La-doped BaTiO$_3$ thin films and studied the composition of the films.

BaTiO$_3$ has a wide range of applications due to its high dielectric constant and significant pyroelectric, piezoelectric and electro-optical effects [11]. Additionally, at high temperatures, barium titanate is paraelectric with a cubic symmetry. When the temperature drops to below 125 °C, paraelectric cubic BaTiO$_3$ transforms to a tetragonal (P4mm), orthorhombic (Amm2) and rombohedral (R3m) structure, thereby exhibiting ferroelectric properties [12-13]. However, this material typically restricts the light absorption due to the broad band gap and insufficient electrical conductivity (the semiconducting behavior) [14]. Currently, these properties can be dramatically improved when it is doped with transition metals or non-metal elements.

A disorder, called Urbach tails or Urbach energy ($E_U$), is one of the most significant parameters that has been extensively investigated for semiconductors, which is a measure of the total disorder occurring in compound such as, thermal, polar, chemical, structural disorder, defects etc. [15-16-17-18]. It should be noted that $E_U$ must remain small. It is estimated from low energy tail states or from the region of the absorption edge where the absorption coefficient shows exponential decay.

In this paper, we set out the relationship between the optical energy gap and the Urbach tail. Besides, we used the density functional theory to accurately report the results regarding the study of the electronic, optical and thermoelectric properties of the pure material and doped by tellurium with various concentrations (2.7%, 4.2% and 8.3%). In this regard, the main idea of our simulation is focused on improving the absorption and electrical conductivity at high temperature for visible-light photoelectrical and thermoelectric applications, respectively.

## 2. THE CALCULATION METHOD

Electronic structures and optical properties of tellurium (Te) doped BaTiO$_3$ are probed by first-principal calculations using the density functional theory implemented in WIEN2k ab initio simulation program [19]. The thermoelectric properties are assessed close to the Fermi level, through the Boltzmann transport theory implemented in BoltzTrap code [20]. We applied the local density approximation (LDA) to describe the exchange-correlation effects and to evaluate the ground state parameters. Tran Balaha modified Becke-Johnson exchange potential (TB-mBJ) is leveraged to exceed the shortcomings in the calculations of electronic properties and band gaps [21]. Static computations are conducted using 500 k- points in reciprocal space. The product of the maximum modulus of reciprocal vector ($K_{max}$) and the smallest of all atomic sphere radii ($R_{min}$) is assumed to be 7 ($R_{min} \times K_{max} = 7$). Self-consistent criterion of the total energy is chosen with a precision of $10^{-6}$ Ry, while the convergence force is fixed to 1mRy/a.u. The other input feedings of the software are selected as $l_{max} = 10$ and $G_{max} = 16$, where $l_{max}$ is the maximum value of the angular momentum vector and $G_{max}$ is the Fourier-expanded charge density. The considered electronic configurations are $5s^2\ 5p^6\ 6s^2$ for Ba, $3s^2\ 3p^6\ 3d^2\ 4s^2$ for Ti, $2s^2\ 2p^4$ for O and $5s^2\ 5p^4$ for Te. The muffin-tin radii (RMT) values are chosen for non-overlapping as follows: 2.42, 1.80, 1.67, and 1.90 (a.u) for each element Ba, Ti, O and Te, respectively.

### 3. RESULTS AND DISCUTION
#### 3. 1. STRUCTURAL PROPERTIES

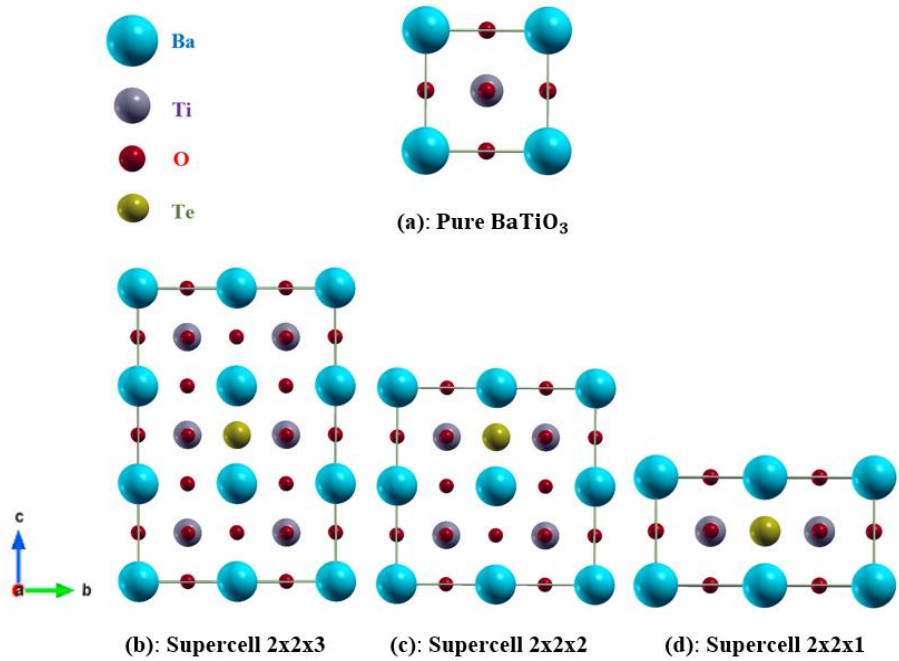

**Fig. 1.** Illustration of schematic views of pure $BaTiO_3$ and doped with tellurium.

The initial cell parameters of the pure structure are a = b = c = 3.9403 Å, which are in great agreement with previous theoretical findings [22] and slightly lower than the experimental measurements ($a_{exp} = b_{exp} = c_{exp} = 3.996$ Å) [23-15]. To study fractional Te substitution, it is necessary to consider a larger cell than the basic unit and to build a periodic fictitious system with the same electronic structure as the real system (see **Fig. 1**). Thus, three kinds of supercells are used 2x2x1 consisting of 20 atoms, 2x2x2 comprising 40 atoms and 2x2x3 containing 60 atoms, which lead to a doping concentration of 8.3%, 4.2% and 2.7%, respectively. The schematic view of the considered configurations is illustrated in **Fig. 1.** More interestingly, All the properties treated in this work (electronic, optical, and thermoelectric properties) are calculated after the full structure relaxation, including lattice constants, atomic positions, and free total energies of supercells in the presence and the absence of impurity, to estimate which oxygen site substitution is energetically more favorable (provide the smallest total energy), which is beneficial for the stability of our structure.

The unit-cell of the pure $BaTiO_3$ showed a cubic symmetry with a space group of Pm-3m (No. 221). Whereas the crystal became a tetragonal structure after doping with P4/mmm symmetry (No. 123) (the space group is distorted). We point out that the arrangement of the atoms in all materials after doping with an element of different features contains imperfections such as disorder, distortions, etc. These deficiencies have profound effects on the physical behavior of systems. Somehow, the incorporation of tellurium atom induces more electrons and may disturb noticeably the symmetry of the lattice due to a mismatch between various parameters, such as ionic radius and electronegativity of $Te^{2-}$ and $O^{2-}$ elements. Fortunately, several energy parameters can be used as criteria for the stability of condensed matters [24]. In this report, the feasibility of synthesis and the relative stability of the doped systems are discussed in the next section by evaluating their defect formation energies.

### 3.2. DOPANT FORMATION ENERGY

To discuss the stability of doped compound, the formation energies are estimated according to the following expression [25]:

$$E_{form}= E\,(BaTiO_3 + Te) - E\,(BaTiO_3) + n_O\mu_O - n_{Te}\mu_{Te} \qquad (1)$$

Where $E\,(BaTiO_3 + Te)$ is the energy of the doped system, $E\,(BaTiO_3)$ denotes the total energy of pure $BaTiO_3$ supercell. $\mu_{Te}$ is the chemical potential per atom of tellurium bulk crystal, deduced from reference material, $\mu_O$ represents the chemical potential of oxygen, which can be calculated using the ground state energy of $O_2$ [26-27], while $n_{Te}$ and $n_O$ are the numbers of added or removed atoms. Namely, the dopant formation energy of Te in the $BaTiO_3$, relates to the energy appropriate to introduce one tellurium atom with chemical potential $\mu_{Te}$, after extracting one oxygen atom from the identical position in the material [25].

**Table 1.** The formation energies $E_{form}$(eV) of pure $BaTiO_3$ and doped systems.

| Concentration of tellurium | 0 % | 2.7 % | 4.2 % | 8.3 % |
|---|---|---|---|---|
| Formation energies | *-16.30 | -24.4 | -13.6 | -1.4 |

**\*The value of the formation energy of pure compound obtained from reference [29]**

The calculated standard formation energies within DFT formalism are summarized in **Table 1**. In fact, the most stable systems are those with smaller formation energies [14]. From the table, it is noted that the formation energies of Te doped $BaTiO_3$ varies according to various concentrations. The tellurium doping amounts of 2.7 % is favorable from the standpoint of formation energy. In the previous case we found theoretically an energy about of -24.4 eV, proving that this system becomes broadly stable than pure $BaTiO_3$. Whereas it is seldom to find such an enhancement in the stability of doped compound as stated in reference [28].

On the other hand, the gradual increase in the concentration of the impurity, leads to a decrease in the stability of the system, as illustrated in **Table 1**. However, we can observe that the calculated of formation energies irrespective of the concentration of tellurium take negative values. Which means the existence of the thermodynamic stability, the compound is energetically favorable and can be conveniently prepared in the experiment.

### 3.3. BAND STRUCTURES AND DENSITY OF STATES

The density of state is one of the most important properties that informs us about the behavior and electronic character of this system. It also allows to know the nature of the chemical bonds between atoms. The total (TDOS) and partial (PDOS) densities of states are calculated at their equilibrium states by LDA + TB-mBJ approach. The projected results between -12 eV and 10 eV are shown in **Fig. 2** and **Fig. 3**. The Fermi level is taken at an energy of 0 eV.

$BaTiO_3$ is a semiconductor of a broad band gap. The local density approximation (LDA) fails to correctly describe the electronic properties by underestimating the band gap value of aforementioned material up to 1.859 eV. However, TB-mBJ yields an indirect band gap value of 2.752 eV, which is consistent with the theoretical calculations already reported [31] and slightly lower than the available experimental results [15-23-30]. In fact, the standard DFT-functional (LDA) reliably describes the exchange–correlation energy, owing to the existence of a derivative discontinuity of this energy with respect the number of electrons [47]. While the success of TB-mBJ functional is due to the introduction of a parameter allowing to modify the relative weight of the two terms in Becke–Johnson potential (for more details Ref. [48]).

In this report, we also illustrate the effect of BaTiO$_3$ doped by tellurium atom (Te) with different concentrations. When the oxygen is superseded by an impurity atom the electronic gap decreased to a different degree depending on the concentration while preserving the semiconducting behavior. As listed in the table below (**Table 2**), TB - mBJ method reproduces band gaps much better than LDA potential and usually gives a substantial value. For 2.7%, 4.2% and 8.3%, it is reduced from 2.752 eV to 1.050 eV, 0.953 eV and 0.500 eV, respectively. These outcomes are largely advantageous for the absorption of visible light. This decrease comes presumably from the interaction of Te − 5p states and Ti − 3d orbitals. In fact, both two adjacent titanium atoms (metal) give an electron to the tellurium atoms (non-metal) forming a stronger and stable polar covalent bond (ΔEN ∼ 0.56) than the mainly ionic and slightly covalent bond (ΔEN ∼ 2) achieved by oxygen in the pure compound. In view of the lessening in the difference in electronegativity ΔEN according to the Pauling scale (for more details see Refs. [32-33]).

**Table 2.** The electronic band gap $E_g$ of pure and tellurium doped BaTiO$_3$ according to different concentrations.

| Compounds | Present work $E_g$ (eV) | | Experimental | Theoretical |
|---|---|---|---|---|
| | LDA | TB-mBJ | | |
| BTO | 1.859 | 2.752 | 3.40 eV [23]<br>3.12 eV [15] | 2.76 eV [31] |
| Te doped BTO: 2.7% | 0.475 | 1.050 | | |
| Te doped BTO: 4.2% | 0.376 | 0.953 | | (--) |
| Te doped BTO: 8.3% | 0.017 | 0.500 | | |

### a) In the case of the pure material

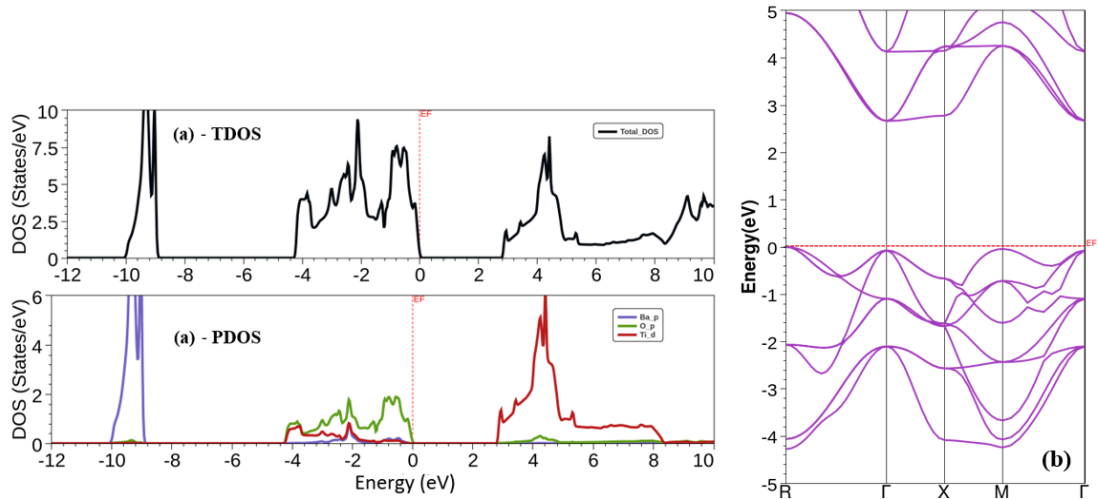

**Fig. 2.** The total and partial density of states of pure $BaTiO_3$ **(a)**. The band structure of undoped $BaTiO_3$ **(b)**. The dashed line indicates the position of the Fermi level.

**Fig. 2 - (a):** The total density of states shows the existence of three regions when using the TB-mBJ approximation. These regions are as follows:

The first region located in the valence band between -12 eV and -9 eV, is principally due to the p states of barium. The second region near the Fermi level comes from the p states of oxygen. Whilst the last region is situated in the conduction band, it is composed essentially of titanium states.

**Fig. 2 - (b):** The band structure for undoped system reveals that the top of the valence band and the bottom of the conduction band are situated at M and $\Gamma$ point of Brillouin zone, respectively. So, the pure $BaTiO_3$ crystal presents an indirect transition. As indicated above, the theoretical indirect gap at M- $\Gamma$ is 2.752 eV. However, the band gap is significantly reduced by incorporating Te element.

**b) In the case of the doped material**

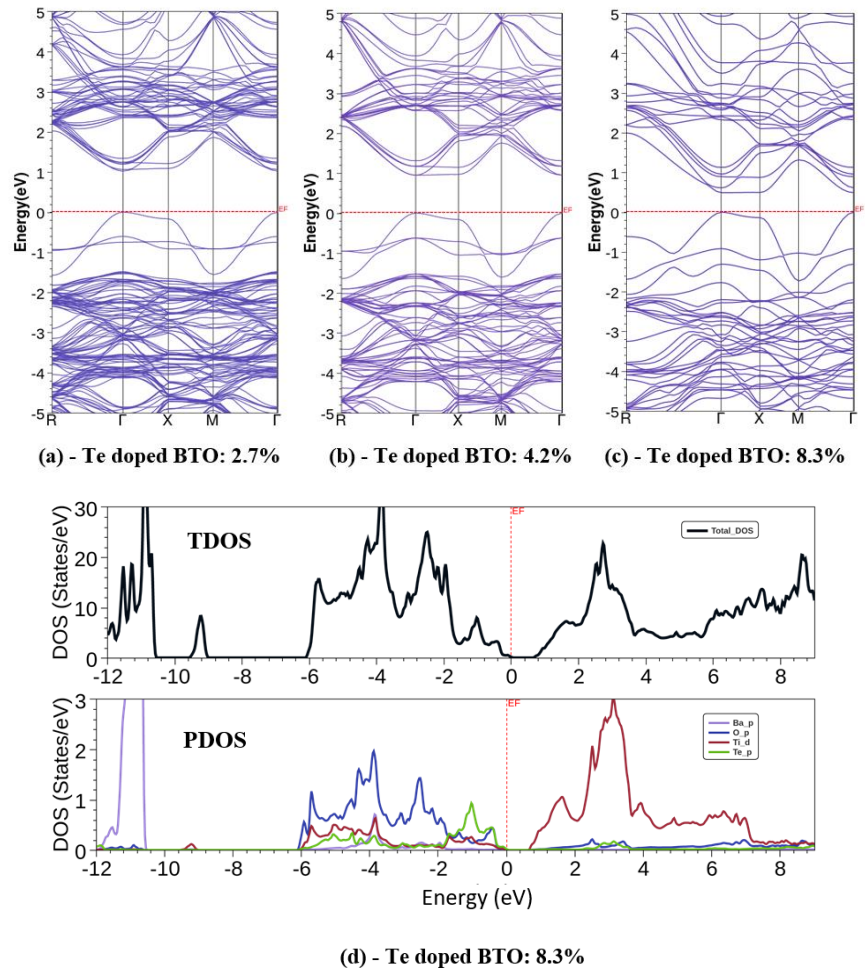

(a) - Te doped BTO: 2.7%    (b) - Te doped BTO: 4.2%    (c) - Te doped BTO: 8.3%

(d) - Te doped BTO: 8.3%

**Fig. 3.** The band structure of $BaTiO_3$ as a function of tellurium concentrations **(a)**, **(b)** and **(c)**. The total and partial density of states of Te doped $BaTiO_3$ at 8.3 % **(d)**.

**Fig. 3 - (d):** The analysis of the variation of TDOS and PDOS close to the Fermi level as a function of the photon energy suggests that the upper part of the valence band arises mainly from $O - 2p$ and $Te - 5p$ states. While a weak contribution of $Ba - p$ and $Ti - 3d$ orbitals is appeared. Whereas the conduction band is predominated by $Ti - 3d$ electrons.

**Fig. 3 - (a) - (b) and (c):** The band structure for doped system indicates that the top of the valence band and the bottom of the conduction band are located at Γ point, proving that the nature of band structure is changed from indirect to direct in doped materials. More interestingly, the lowest energy required for electron excitation is defined by band gaps in insulators and semiconductors. However, it remains unclear whether the photons will be absorbed by material or not, as the indirect band gap exhibits a low optical transition due to finite momentum [49]. Hence, the enhancement of electronic band gap in $BaTiO_3$ after doping

indicates the significant improvements of conductivity of compound due to the convenience of hole electron recombination in direct transitions. Fermi level is shifted towards the valance band edge suggesting that all the doped compounds are p-type of materials, which is confirmed in 'THERMOELECTRIC' section.

### 3.4. OPTICAL STUDIES

The study of the optical properties of solids proved to be a powerful tool for understanding the electronic properties of materials. In this part the study of optical properties is carried out using TB-mBJ approximation, due to its success in the determination of the optical band gap with appreciable accuracy.

The optical properties of any material can be described by the complex dielectric function [34]:

$$\varepsilon(\omega) = \varepsilon_1(\omega) + i\,\varepsilon_2(\omega) \qquad (2)$$

$\varepsilon_1(\omega)$ and $\varepsilon_2(\omega)$ denote the real and imaginary parts of the dielectric function, respectively. $\varepsilon_2(\omega)$ is directly related to the electronic band structure. It is calculated by summing all transitions from occupied states to unoccupied states using the well-known relation [34]:

$$\mathrm{Im}\,\varepsilon(\omega) = \varepsilon_2(\omega) = \frac{4\pi^2 e^2}{m^2 \omega^2} \sum_{i,j} \int |\langle i|M|j\rangle|^2 - (f_i(1-f_i))\delta(E_f - E_i - \hbar\omega) d^3k \qquad (3)$$

Where M is the dipole matrix, i and j are the initial and final states, respectively. $f_i$ is the distribution of Fermi as a function of $i - $ th state, and $E_i$ is the energy of electron in the $i - $ th state.

The real part of the dielectric function can be obtained using the Kramers–Kronig relation [34]:

$$\varepsilon_1(\omega) = 1 + \frac{2}{\pi} P \int_0^\infty \frac{\varepsilon_2(\alpha)\alpha d\alpha}{\alpha^2 - \omega^2} \qquad (4)$$

Where P represents the main value of the integral.

The absorption coefficient is obtained with the subsequent relation [35]:

$$\alpha(\omega) = \frac{\sqrt{2}}{c} \omega \sqrt{-\varepsilon_1(\omega) + \sqrt{\varepsilon_1(\omega)^2 + \varepsilon_2(\omega)^2}} \qquad (5)$$

The optical conductivity is given from reference [35]:

$$\text{Re } \sigma_{\alpha\beta}(\omega) = \frac{\omega}{4\pi} \text{Im } \varepsilon_{\alpha\beta}(\omega) \tag{6}$$

### 3.4.1. DIELECTRIC FUNCTION

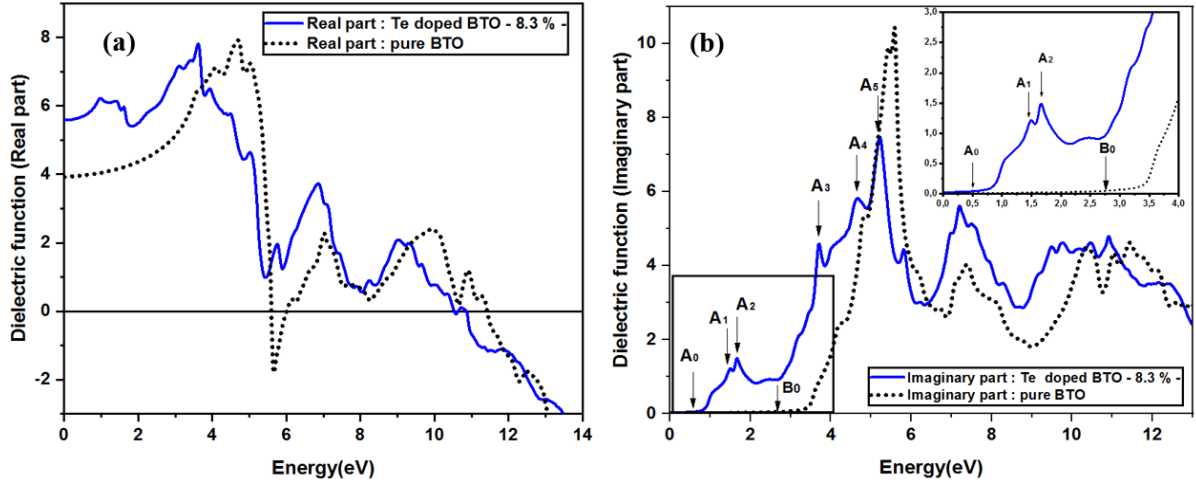

**Fig. 4. (a)** The real part of the dielectric function of pure and Te doped BaTiO$_3$ at 8.3 %. **(b)** The imaginary part of the dielectric function of pure and Te doped BaTiO$_3$ at 8.3 %.

It is well known that the imaginary part of the dielectric function $\varepsilon_2(\omega)$ is closely related to the electronic band structure and exhibits the light absorption in the material. Whereas the real part of the dielectric function $\varepsilon_1(\omega)$ refers to the dispersion of the incident photons by crystals [36]. **Fig. 4 - (b)** shows the imaginary part of the dielectric function. One can notice that the curve of the pure BaTiO$_3$ indicates an energy about of 2.760 eV at B$_0$ point (see **Fig. 4 - (b)**). This threshold value comes mostly from the electronic transitions between the valence band maxima and the conduction band gap minima. Usually, it is referred to a band gap energy. After doping, the peaks appeared in the region between 0.50 eV and 5.22 eV at 8.3 %, which corresponding to a significant absorption of our doped compound as studied below.

Several peaks occurring in $\varepsilon_1(\omega)$ spectra, indicate plasmonic resonance and polarization arising inside the material when radiations of a sufficient frequency interact with its surface (**Fig. 4 – (a)**) [50]. The static values of $\varepsilon_1(\omega)$ at zero frequency or energy, which are represented as $\varepsilon_1(0)$ are extracted directly from the computed spectra. $\varepsilon_1(0)$ rises with Te doping ( $\varepsilon_1(0)$ = 5.55) compared to pure BaTiO$_3$ ($\varepsilon_1(0)$ = 3.98), while the optical band gap decreases as discussed in the sections below. Hence, static values obey the relation $\varepsilon_1(0) \approx 1 + (\hbar\omega_p + E_g)^2$ that illustrates Penn's model [51]. Where, $\hbar$, $\omega_p$ and $E_g$ express Plank's constant, plasma frequency and optical band gap, respectively. In fact, by growing Te content, concentration of carriers is increased, and the nucleus exhibits a small grip on the electrons of the valence shell, which improves noticeably the dielectric constant (see Ref. [50]). The computed $\varepsilon_1(\omega)$ increases from its static value and reaches maximum values at 4.4 eV and 4 eV for $BaTiO_3$ and Te - doped $BaTiO_3$ at 8.3%, where it exhibits plasmonic resonance [52]. Obviously, the peaks of maximum intensity are shifted towards lower energy region meaning that the light through Te - doped $BaTiO_3$ disperses more than $BaTiO_3$ because of longer wavelength [53]. In contrast, peaks of varying intensities appearing in the vicinity of the maximum peak are also generated due to the resonance of free carriers' mobility, which are an immediate consequence of doping process [51]. $\varepsilon_1(\omega)$ of pure $BaTiO_3$ starts to decrease continuously and turns negative before the energy value of 6 eV. After showing some features, the real part of dielectric function of Te - doped $BaTiO_3$ attains the negative values at an energy of 11 eV. Such a negative $\varepsilon_1(\omega)$ indicates that reflecting nature of the studied materials is similar to a metallic compound that reflects all the light falling on it (the surface entirely reflects the impinging light) [51]. Hence, doping with chalcogen element (Te) significantly narrows width of the metallic region.

### 3.4.2. OPTICAL ABSORPTION COEFFICIENT AND OPTICAL CONDUCTIVITY

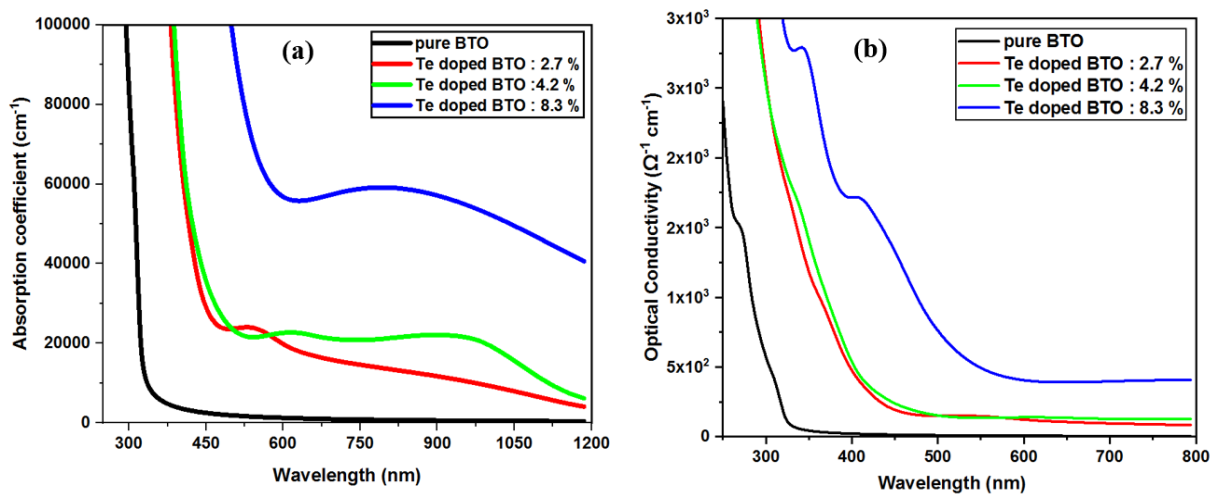

**Fig. 5.** **(a)** The absorption coefficient of pure and tellurium doped $BaTiO_3$ as a function of wavelength. **(b)** The optical conductivity as a function of wavelength.

The optical absorption spectrum of semiconductor materials exhibits an extensive prerequisite to describe their composition and optical band gap. Moreover, optical absorption spectra of semiconducting compounds are subdivided into three main ranges. The former stems from defects and impurities (weak absorption). The last provides the optical energy gap (strong absorption). While the middle region occurs due to perturbation of structural and disorder of the system (for more details see Refs. [16]). The results obtained suggest that the shape of the absorption curve depends heavily on the concentration of the dopant introduced. The absorption ability is enhanced for tellurium doped $BaTiO_3$ compared to the undoped compound in the visible light (380~790 nm). **Fig. 5 - (a)** shows the optical absorption spectrum as a function of the wavelength. One can obviously see that the pure $BaTiO_3$ has a broad absorption coefficient in the UV (ultraviolet) range around 300000 cm$^{-1}$ at 250 nm and decreases rapidly until it abrogates in the visible and IR (infrared) games. Therefore, such ranges can be regarded as transparent. The absorption behavior of the pure compound is in accordance with the previous studies [37-11] and experimental measurement [15]. In contrast, the absorption edge extends into the visible light range when $BaTiO_3$ is doped with Te element. At 2.7% and 4.2%, we observe an average absorption intensity in the region between λ = 390 nm and λ = 1200 nm (visible and near IR). While a widespread absorption appears at 8.3%. It provides evidence that the ranges of $BaTiO_3$ activities after doping became significant in the visible region. In fact, the intensity and sharpness of the peaks in absorption spectrum are strongly dependent on the number of occupied and unoccupied electronic states in the valence and conduction bands, respectively, as well as on the energy level degeneracy [54]. On the other hand, absorption spectrum shifts towards longer wavelengths due to different allowed direct optical transitions from filled to empty states, and due to a large carrier's concentration with moderately large mobility in the Te-doped $BaTiO_3$, that occurs by growing Te-concentration [55]. Moreover, absorption ability is squarely related to $E_g$. Hence, the main factor of a good performance in the absorption of visible light is due to the photon energy that is extremely higher than the band gap (hν > $E_g$), the thing that makes the photons widely absorbed. The absorption edges observed at the longer wavelengths (shorter energy) define the limit of the band gap for each compound (hν = $E_g$) and yield a sharp decrease of absorption ability. Often, the observed behavior occurs when band gap of the material is greater than excited photon energy (hν < $E_g$), therefore, in this range optical transitions are not permitted [56]. Our findings suggest that the band gap of pure $BaTiO_3$ (~2.752 eV) is significantly reduced by increasing Te concentration (~1.050 eV for BTOTe: 2.7%, ~0.953 eV for BTOTe: 4.2% and ~0.500 eV for BTOTe: 8.3%). Each value of $E_g$ corresponds to a specific wavelength, which might explain the observed shift in the sharp decreases. These results are suitable for photoelectrical applications.

**Fig. 5 - (b)**. The plot shows the optical conductivity as a function of wavelength. Namely, the optical conductivity is strictly related to the absorption spectra [35]. For the pure compound, it is clearly noteworthy that the conductivity vanishes in the visible region. Whereas it increases gradually after doping until the appearance of critical points at 8.3% related the optical transitions.

### 3.4.3. OPTICAL BAND GAP DETERMINATION

The relation between the absorption coefficient α, the incident photon energy hν and the optical energy gap $E_g$ can be established by using Tauc's relationship mentioned above [34]:

$$(\alpha h\nu)^m = A_0(h\nu - E_g) \qquad (7)$$

Where $A_0$ is a constant related to the transition probability [34] and in some cases referred to as a band tailing parameter. The constant m is usually designated as the power factor of the transition mode which depends upon the nature of materials and photon transition [16]. In general terms, the value of the exponent indicates the type of electronic transitions, whether allowed or forbidden and whether direct or indirect [16]. Typically, for semiconductors, the permitted transitions predominate the absorption processes giving either m = 1/2 for indirect transitions or m = 2 for direct transitions as reported by Brian D. Viezbicke et al, [57] and confirmed by Jennifer B. Coulter et al, [58]. Although, the pure barium titanate has an indirect gap, the value of the exponent m is 1/2. While Te-doped $BaTiO_3$ at 2.7%, 4.2% and 8.3% it is equal to 2 because the electronic band nature becomes direct [38]. The obtained outcomes may provide guidance for experimental research.

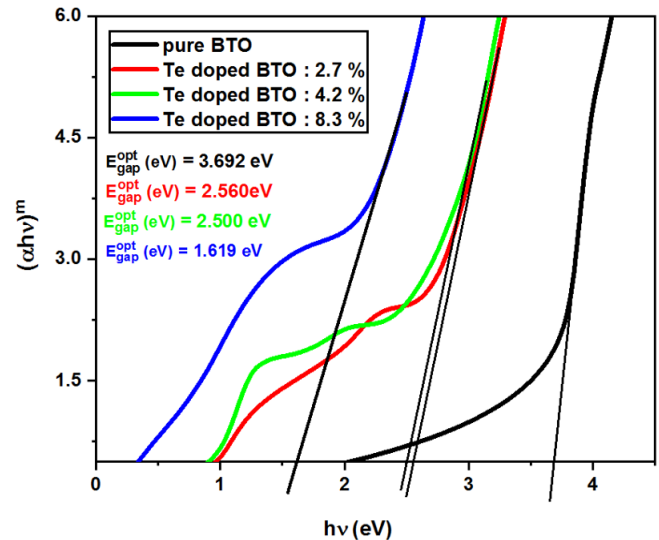

**Fig. 6.** Plot of $(\alpha h\nu)^m = f(h\nu)$ for the estimation of the optical energy gap. The value of the exponent m is 1/2 for the pure compound (indirect gap: $(\alpha h\nu)^{1/2}\ 10^2$ ) and 2 for tellurium doped barium titanate (direct gap:$(\alpha h\nu)^2\ 10^{11}$).

As illustrated in **Fig. 6**, the optical band gap $E_g^{opt}$ for various concentrations can be acquired by plotting $(\alpha h\nu)^m = f(h\nu)$ [38]. It indicates that the achieved plotting forms a straight line in a specific area. To determine the value of the optical energy gap, we extended this straight line to intercept (hν)-axis at $(\alpha h\nu)^m = 0$ [16]. The expected values of the energy gap are tallied in **Table 3**. It is found that the optical band gap decreased with increasing the Te concentration in compound.

**Table. 3.** The optical band gap and Urbach energies values of $BaTiO_3$ as a function of tellurium concentrations.

| Compounds | Urbach energy $E_U$ (eV) | Optical band gap $E_g^{opt}$ (eV) | | |
|---|---|---|---|---|
| | | Present work | Theoretical | Experimental |
| BTO | 0.0918 | 3.692 | 2.640 [41] | 3.20 [42] |
| Te doped BTO: 2.7% | 0.3010 | 2.560 | | |
| Te doped BTO: 4.2% | 0.3270 | 2.500 | (--) | |
| Te doped BTO: 8.3% | 0.5010 | 1.619 | | |

### 3.4.4. URBACH ENERGY OR URBACH TAIL

Throughout the absorption coefficient curve, Urbach tail exhibits the exponential part existing close to the optical band edge [16]. Mostly, such tails can originate from structural, chemical, and thermal defects [15]. It should be noted that the Urbach energy $E_U$ supplies information on the various types of disorders existing in the material and faintly depends on temperature (see Ref. [39]). The energy of the band tail is related squarely to the absorption coefficient by the Urbach empirical rule. Which is given from references [16-39]:

$$\alpha = \alpha_0 e^{\left(\frac{h\nu}{E_U}\right)} \qquad (8)$$

Where $\alpha_0$ is a constant. It is well known that $E_U$ can be reached from the inverse of straight-line slope of plotting $\ln(\alpha)$ versus the incident photon energy hν.

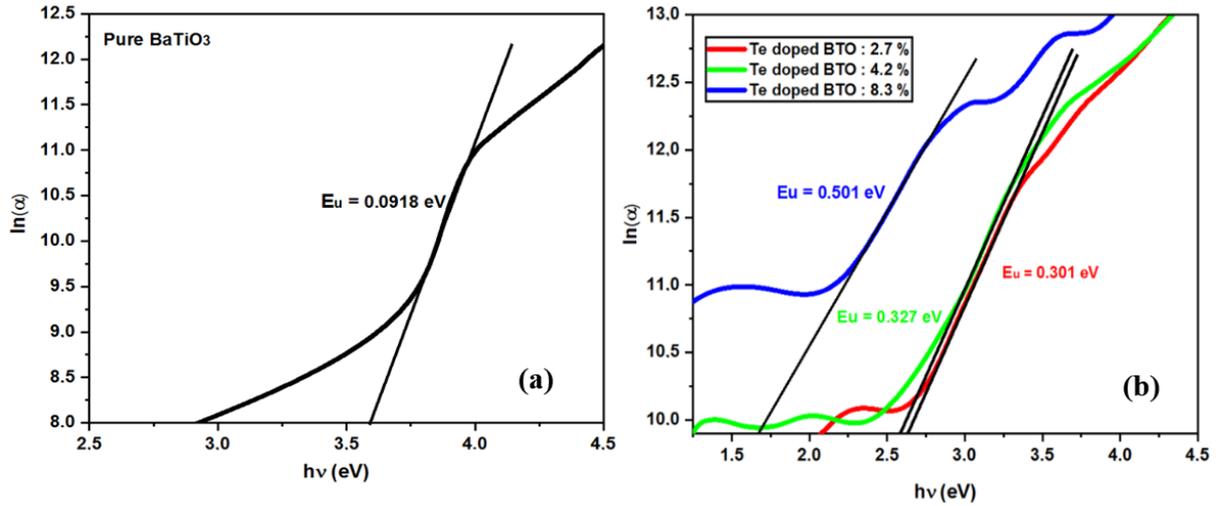

**Fig. 7.** Plot of $\ln(\alpha) = f(h\nu)$ for pure **(a)** and Te doped $BaTiO_3$ **(b)**, from which the Urbach energy can be obtained.

**Fig. 7 - (a)** and **(b)** illustrate the variation of $\ln(\alpha)$ against $h\nu$. It can be noted that the values of the Urbach energy $E_U$ increased with increasing the concentration of the tellurium in the $BaTiO_3$ compound (the worsening of disorder in the material). Whereas it varies inversely to the behavior of the optical band gap as tabulated in **Table 3.**

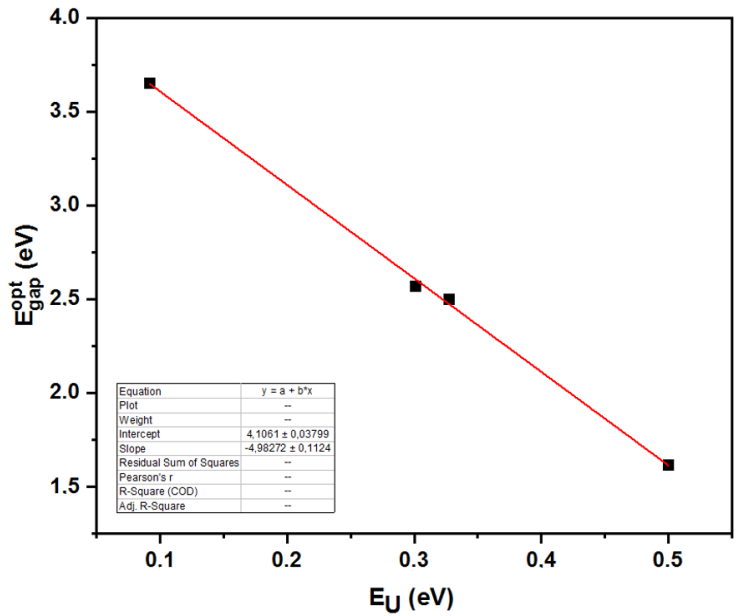

**Fig. 8.** The relation between the band gap $E_g^{opt}$ and Urbach tail $E_U$.

It can be noted that the mutual relationship between the optical energy gap and the Urbach tail can be tailored by the linear regression as shown in **Fig. 8**, to yield the theoretical equation as mentioned below:

$$E_g^{opt} = 4.1061 - 4.98272\, E_U \tag{9}$$

The above linear relation may be useful for theoretically approximating the value of the optical band gap in the interval of the Urbach tail. We emphasize that this concept is adopted for other semiconductors [16-17-40].

### 3.5. THERMOELECTRIC STUDIES

To get an insight into the thermoelectric behavior of compounds, the transport coefficients are determined close to the Fermi levels as a function of temperatures, such as Seebeck coefficient (S), electrical conductivities relative to relaxation time ($\sigma/\tau$) and the figure of merit (ZT). The thermoelectric (TE) properties of materials can be determined close to the Fermi level, by using the Boltzmann transport theory implemented in the BoltzTrap code [20]. Further, the thermoelectric materials can be widely used to convert heat to electrical power. The efficient of this conversion is determined by the dimensionless figure of merit (ZT), which is defined by the mathematical expression $ZT = \sigma S^2 T/\kappa$ [23]. S, T, $\sigma$, and $\kappa$ stand for Seebeck coefficient, absolute temperature, electrical and thermal conductivities, respectively.

In Boltzmann equations, the transport coefficient of electrical conduction can be determined as [43]:

$$\sigma_{\alpha\beta} = \frac{1}{N}\sum_{i,k} \sigma_{\alpha\beta}(i,k) \frac{\delta(\varepsilon - \varepsilon_{i,k})}{\delta(\varepsilon)} \tag{10}$$

$$\sigma_{\alpha\beta}(i,k) = e^2 \tau_{i,k} v_\alpha(i,\vec{k}) v_\beta(i,\vec{k}) \tag{11}$$

In the above relations, N represents the number of k-points, $\sigma_{\alpha\beta}(i,k)$ denotes the transport tensor that is written as a function of relaxation time $\tau$ and group velocity $v$, while $e$ is the electron charge.

The electrical conductivity $\sigma$ and Seebeck coefficient S, can be written as a function of chemical potential and absolute temperature, in the form of distribution function as developed through the following equations [43-44-25]:

$$\sigma_{\alpha\beta}(\alpha,\mu) = \frac{1}{\Omega}\int \sigma_{\alpha\beta}(\varepsilon)\left[-\frac{\partial f_0(T,\varepsilon,\mu)}{\partial \varepsilon}\right]d\varepsilon \tag{12}$$

$$S_{\alpha\beta}(T,\mu) = \frac{1}{eT\Omega\sigma_{\alpha\beta}(T,\mu)}\int \sigma_{\alpha\beta}(\varepsilon)(\varepsilon - \mu)\left[-\frac{\partial f_0(T,\varepsilon,\mu)}{\partial \varepsilon}\right]d\varepsilon \tag{13}$$

The tensor indices α and $\beta$ stand for the wave vector, and the electronic charge, respectively. Whereas Ω is the volume of the unit cell, and $f_0$ is the Fermi-Dirac distribution function. The electronic thermal conductivity of $BaTiO_3$ and systems can be determined as [44]:

$$\kappa_e = \frac{1}{e^2 T} \int (\varepsilon - \mu)^2 \, \sigma(\varepsilon) \frac{\partial f(\varepsilon)}{\partial \varepsilon} \, d\varepsilon \tag{14}$$

μ is the Fermi energy.

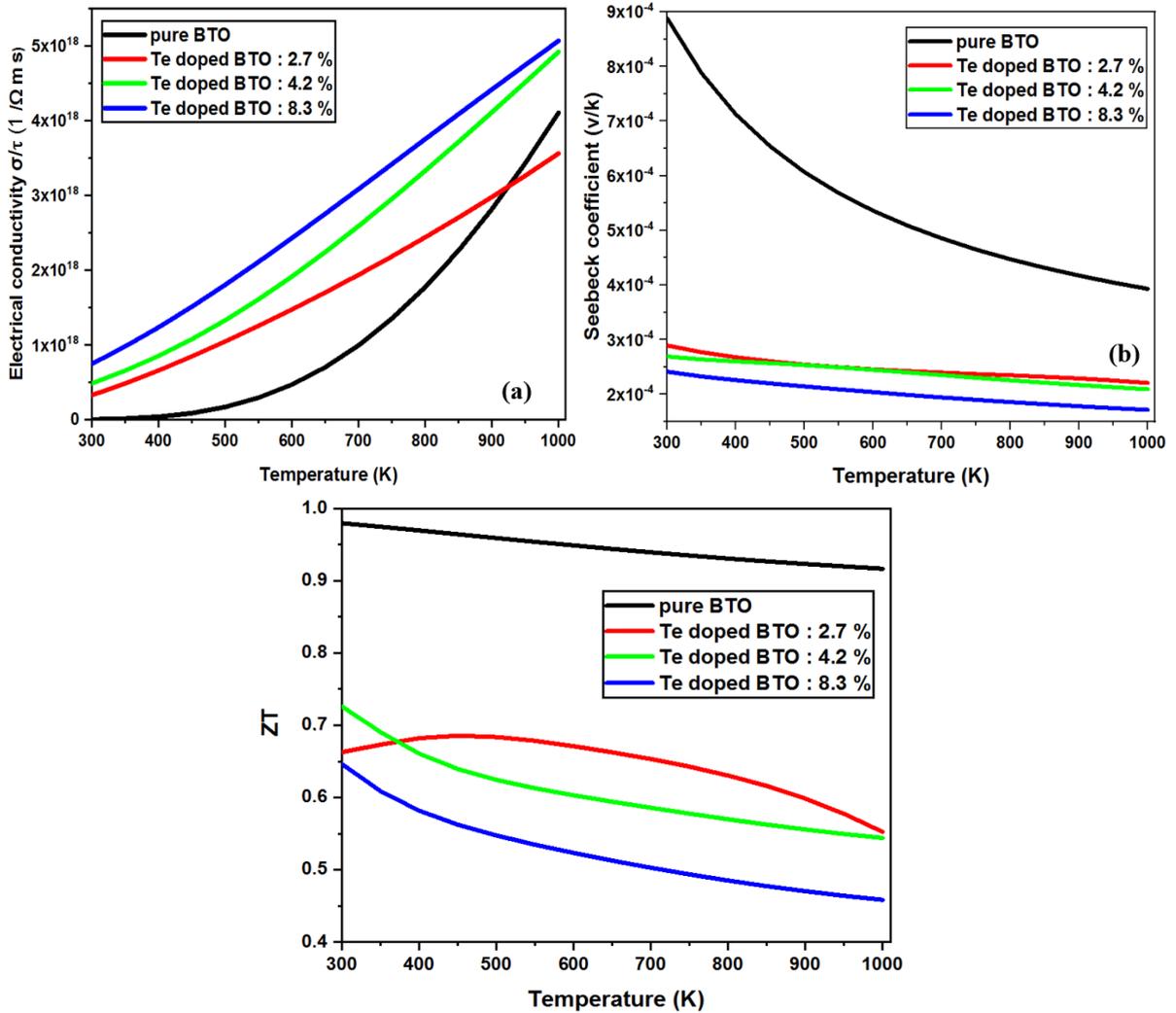

**Fig. 9.** The traces of electrical conductivity **(a)**, Seebeck coefficient **(b)** and figure of merit ZT **(c)** depending on temperature, according to different concentrations.

As depicted in **Fig. 9 - (b)**, the positive values of the Seebeck coefficient (S) for the pure and Te doped compound in the investigated temperature zone, reflect the p-type semiconductor. In **Fig. 9 - (a)**, the plot shows electrical conductivity as a function of temperature. It obviously appears from these results that (σ/τ) has the direct proportionality with temperature and becomes more substantial according to the gradual increase of tellurium concentration in compound. This proportional reveals that there are greater concentrations of carriers and also, it means that the electrons are thermally excited to the conduction bands. In fact, with the increase in temperature the carriers are excited to cross band gap and cause continuously increasing electrical conductivity.

**Fig. 9 - (c)** illustrates the thermoelectric figure of merit (ZT). It is a dimensionless quantity used to measure the thermoelectric efficiency of materials. It is related to several parameters such as Seebeck coefficient, temperature, electrical conductivity, and thermal conductivity. Greater the value, higher will be the thermoelectric efficiency. It increases with increasing electrical conductivity and Seebeck coefficient while decreases with increasing thermal conductivity, as stated in reference [45].

In the present research, one can notice that the maximum value of ZT is found approach to the unity for the pure material at 300 K. Which is consistent with the previous studies [46]. Whilst it reaches 0.66, 0.73 and 0.65 at 2.7%, 4.2% and 8.3 % respectively. We estimate that the observed decrease is attributed to the increase in thermal conductivity. This result can still be selected as a suitable candidate for thermoelectric applications at high temperature.

# CONCLUSION

In summary, the formation energy, electronic, optical, and thermoelectric properties of $BaTiO_3$ are studied by first-principles calculations based on density functional and the semi-classical Boltzmann theories. The band gap obtained by LDA approximation, is adjusted by mBJ method. It is found from first principles results that the introduction of tellurium atom into the lattice decrease efficiently the electronic gap from 2.75 eV to 1.030 eV at 2.7%, 0.953 eV at 4.2% and 0.500 eV at 8.3%. Accordingly, the range of activity of barium titanate after doping became extensive in the visible region. The transformation in resonance peak to lower energy region is observed by increasing Te - concentration. Additionally, Urbach energy varied inversely to the optical energy gap, the linear relationship between these parameters is established. Moreover, the thermoelectric properties of the pure and doped compound are estimated close to the Fermi levels as a function of temperatures. The positive values of the Seebeck coefficient shows a p-type behavior. The incorporation of Te atom yields an increase in the electrical conductivity of the $BaTiO_3$ compound. Therefore, the achieved maximum figure of merit (ZT) is 0.66 at 2.7%, 0.73 at 4.2% and 0.65 at 8.3%. Our theoretical results need to be confirmed by experiments and can be beneficial for thermoelectric and photoelectric applications.

**Acknowledgments**

The authors are thankful to Prof. P. Blaha and Prof. K. Schwarz at Wien Technical University for the Wien2k package and the group of WIEN2K for useful discussions.